\RequirePackage{fix-cm}

\documentclass[smallextended]{svjour3}
\smartqed
\usepackage{graphicx}
\usepackage{dcolumn}
\usepackage{bm}

\def\iota{\imath}
\date{\today}
\begin{document}
\title{Relativistic and non-relativistic quantum Brownian motion in an anisotropic dissipative medium}

\author{Ehsan Amooghorban\and Fardin Kheirandish}

\institute{E. Amooghorban \at
              Department of Physics, Faculty of Basic Sciences,
Shahrekord University, P.O. Box 115, Shahrekord 88186-34141, Iran \\
                           \email{Ehsan.amooghorban@sci.sku.ac.ir}           
           \and
           F. Kheirandish \at
              Department of Physics, Faculty of Science, University
of Isfahan, Hezar Jarib Ave., Isfahan, Iran
}

\maketitle

\begin{abstract}
Using a minimal-coupling-scheme we investigate the quantum Brownian motion of a particle in an
anisotropic-dissipative-medium under the influence of an arbitrary potential in both relativistic and
non-relativistic regimes. A general quantum Langevin equation is derived and explicit expressions for
quantum-noise and dynamical variables of the system are obtained. The equations of motion for the canonical
variables are solved explicitly and an expression for the radiation-reaction of a charged particle in the
presence of a dissipative-medium is obtained. Some examples are given to elucidate the applicability of this
approach.
\end{abstract}\\

PACS numbers:{\,03.70.+k , 11.10.Ef, 05.40.Jc, 05.40.Ca}
\keywords{Anisotropic dissipative medium, Langevin-equation, Coupling tensor, Radiation reaction, Dirac equation, Cherenkov radiation.}

\section{Introduction}
In the present paper we introduce a scheme for investigating the
quantum dynamics of a particle embedded in an anisotropic
dissipative medium, under the influence of an arbitrary potential.
This problem is fundamental to many fields of physics: statistical
mechanics, chemical physics \cite{1}- \cite{9}, condensed matter
\cite{10}-\cite{12}, quantum optics \cite{13}-\cite{18a}, quantum
information and atomic physics \cite{19}. Our purpose is to show
that our approach is a macroscopic description that can be applied in a general way from the classical to the relativistic domain and is consistent with physics requirements, in particular causality and fluctuation-dissipation theorem.

For investigating the quantum mechanical description of a dissipative system, there are usually two approaches:
the first approach is based on the assumption that the damping of the system is caused by an irreversible
transfer of its energy to the reservoir due to the coupling of the system with the reservoir \cite{20}- \cite{22}. Therefore, the loss of energy is phenomenologically described in terms of a frictional force. In addition, the system also is subject to a fluctuating or noise force and that dissipation and fluctuations are related.
The second approach is essentially a rigorous one in which the effect of dissipation is introduced by ingeniously construction a suitable Lagrangian or Hamiltonian for the system
\cite{25,26}. Historically, the first Hamiltonian was introduced by Caldirola \cite{27} and rederived independently by Kanai \cite{28} and afterward by several others \cite{29,30}. They employed a time dependent mass in such
a way that a friction term appears in the corresponding equation of motion. There are significant difficulties in the quantum mechanical solution of the Caldirola-Kanai Hamiltonian, for example quantizing in this way violates the
uncertainty relation or canonical commutation rules and the former vanishes as time tends to infinity
\cite{31}-\cite{34}.

These common approaches to quantum dissipative systems can generally be divided into two classes. Working in Schrodinger picture and Heisenberg picture, that the dynamics is described in terms of generalized master
equations \cite{35,36} and Langevin equations \cite{37,38}, respectively. Although, the effects of environmental degrees of freedom on the system can be investigated with the method of Feynman-Vernon influence functional by
integrating out environment variables within the context of the closed-time-path formalism \cite{39}-\cite{46}.
The more complicated interaction by considering nonlinear couplings of the particle with the reservoir has also been studied in \cite{23,24}. Furthermore, the relativistic Brownian motion has also been discussed in \cite{47,48}.

The main purpose of the present work is to develop a canonical theory of Brownian motion to extract the classical, the nonrelativistic and the relativistic quantum Langevin equation consistent with fluctuation-dissipation theorem. The result is a relation between noise correlations and susceptibility in frequency domain where the proportionality constant depends on temperature. To achieve this goal, we first introduce an appropriate Lagrangian to
including the dissipative effect in a consistent form and then generalize this Lagrangian to charged particles in presence of the electromagnetic fields. This prepares not only the grounds to survey the radiation reaction but extracts a Dirac  equation for a relativistic particle.  Our approach suggests the simplest way in which the dissipation and the fluctuation effects can emerge from the classical to the relativistic quantum theory. On this base, it is enough using a minimal coupling scheme to obtain a suitable Hamiltonian and also the motion equations to describe the dissipative system.

The layout of the paper is as follows: In Sec. 2, a Lagrangian for the total system is proposed and a classical
treatment of the damped system is investigated. In Sec. 3, we use the Lagrangian introduced in the Sec. 2 to
canonically quantize the system and obtain the corresponding Langevin equation. Subsequently in Sec. 4, as a
simple application, we calculate the spontaneous decay rate of an initially excited two-level atom embedded in an
anisotropic dissipative medium then this formalism is generalized to describe the radiation reaction of a
charged particle in this medium. In Sec. 4, a modified Lagrangian is introduced to describe the relativistic effects of  charged particles embedded in an anisotropic dissipative medium. Then, Cherenkov radiation that emitted by medium due to the motion of a charged particle moving through which is examined. Finally, conclusions are given in Sec. 5.

\section{Classical dynamics}
Classical and quantum description of a dissipative system under influence of a potential
$V$ can be accomplished by modeling it which is in interaction with a heat bath. We assume that the heat bath consists of a continuum of three dimensional harmonic oscillators labeled by a
continuous parameter $\omega$. This kind of heat bath could be a
model for an elastic solid, a dissipative medium and electromagnetic
field that known as the Hopfield model \cite{49}. This model can
be also used to describe the quantization of electromagnetic field
in presence of an amplifying magnetodielectric medium~\cite{50} where the electric and
magnetic properties of the medium are modeled by two independent sets of
harmonic oscillators. In order to examine the classical and the quantum treatment of a linear dissipative system,
we start the following classical Lagrangian for the total system
\begin{equation} \label{a1}
L(t) = L_{\rm m}  + L_{\rm s} +L_{\rm int}.
\end{equation}
The first term $L_{\rm m}$ is the Lagrangian of the medium which is a continuum of three dimensional harmonic
oscillators defined by
\begin{eqnarray} \label{a2}
L_{\rm m} \, = \frac{1}{2}\int_0^\infty  d\omega\, [{\bf \dot X}^2(\omega,t)-
\omega ^2 {\bf X}^2(\omega,t)],
\end{eqnarray}
where ${\bf X}(\omega,t)$ is the dynamical variable of the medium. The second term $L_{\rm s}$ is the Lagrangian of
the main system with mass $m$, position ${\bf q}$ and potential $V(\bf q)$
\begin{equation} \label{a3}
L_{\rm s} \, = \frac{1}{2}m\,{\bf \dot q}^2(t) -
V(\bf q) ,   \\
\end{equation}
and $ L_{\rm int} $ is the interaction term defined by
\begin{eqnarray} \label{a4}
\hspace{-1cm} L_{\rm int}  = \int_0^\infty {d\omega \,}f_{ij} (\omega )\,\dot { q}_i (t)\,{X}_{i}(\omega,t),
\end{eqnarray}
where summation should be done over the repeated indices and $f_{ij}(\omega )$ is the coupling tensor which for
an isotropic medium is written as $f(\omega)\delta_{ij}$. We can simply obtain the classical equations of motion
from the Euler-Lagrange equations
\begin{eqnarray}\label{a5}
&& \frac{d}{dt}\left(\frac{\delta L}{\delta \dot q_i(t)}\right) -
\frac{\delta L}{\delta
q_i(t)} = 0\hspace{+2cm}i=1,2,3  \nonumber\\
&& m{\rm \ddot {\bf q}}(t) + {\bf \nabla}_q V({\bf q}) = -
\dot {\bf {R}}(t),
\end{eqnarray}
and
\begin{eqnarray}\label{a6}
&& \frac{d}{dt}\left(\frac{\delta L}{\delta
\dot{X}_i(\omega,t)}\right) - \frac{\delta L}{\delta X_
i(\omega,t)} = 0 \hspace{+2cm}i=1,2,3 \nonumber\\\nonumber\\
&& \ddot {X}_i(\omega,t) + \omega ^2\, X_i(\omega,t)  =
\dot{q}_j (t)
f_{ji}(\omega ),
\end{eqnarray}
where the components of the field ${\bf R}$ is defined by
\begin{equation}\label{a7}
R_i = \int_0^\infty{d\omega } f_{ij} (\omega )X_j(\omega).
\end{equation}
The formal solution of the field equation (\ref{a6}) is
\begin{equation} \label{a8}
{ X}_i (\omega,t )   =  { \dot{X}}_i (\omega,0 ) \frac{{\sin \omega
t}}{\omega } +
 { X_i} (\omega,0 )\cos \omega t + \int_0^t  {dt'}
 \frac{{\sin \omega (t - t')}}{\omega }
 { f_{ji}}(\omega) { \dot q}_j(t^\prime)
\end{equation}
where the first term is the inhomogeneous solution of Eq. (\ref{a6})
and the second term is the homogeneous one. We will show
later, the homogeneous solution after quantization becomes a
noise operator. Now by substituting Eq.~(\ref{a8}) in the integrand of Eq.~(\ref{a7}), we find that the field $\bf R$ is as follows
\begin{equation}\label{a9}
{{ R}_i}(t) = \int_0^\infty  {dt'\,} \chi_{ij}(t - t'){\dot {
q}}_j(t')\, + { R}_{i}^N(t),
\end{equation}
where $\chi_{ij}$ is the causal susceptibility tensor of the medium and in terms of the coupling tensor $
f_{ij}$ can be written as
\begin{eqnarray} \label{a10}
{\chi }_{ij}(t) = \int_0^\infty
{d\omega } \frac{{\sin \omega t}}{\omega }{ f}_{il} {f}_{jl}
(\omega ){\rm \Theta(t)},
\end{eqnarray}
where ${\rm \Theta(t)}$ is the Theta function.
The second term in Eq. (\ref{a9}) is a noise function which
in terms of the coupling tensor $f_{ij}$ are obtained as
\begin{equation} \label{a11}
{R}_{i}^N(t) = \int_0^\infty  {d\omega } \,f_{ij}(\omega )\, \left(
{{\dot{X}}_j(\omega,0)\frac{{\sin \omega t}}{\omega } + { X}_j(\omega,0)
\cos \omega t} \right).
\end{equation}
It is easily shown that Eq.~(\ref{a10}) is the origin of the significant Kramers-Kronig relations
\begin{equation}\label{a11/1}
{\rm Re}[ \chi _{ij} (\omega )] = P\int_{ - \infty }^\infty  {\frac{{d\nu }}{\pi }}
\frac{{{\mathop{\rm Im}\nolimits}[ \chi _{ij} (\nu )]}}{{\nu  - \omega }},\,\,\,\,\,\,\,\,\,\,\,{\rm Im}[ \chi _{ij} (\omega )] =  - P\int_{ - \infty }^\infty {\frac{{d\nu }}{\pi }} \frac{{{\mathop{
\rm Re}\nolimits} [\chi _{ij} (\nu )]}}{{\nu  - \omega }}.
\end{equation}
Here, the symbol $P$ denotes the Cushy principal value and
\begin{equation}\label{a12}
{\chi}_{ij} (\omega) =\frac{1}{\sqrt{2 \pi}} \int_{-\infty}^\infty {dt
}{\chi }_{ij}(t) e^{\imath\omega t}.\nonumber
\end{equation}
Note that, for a definite susceptibility tensor $\chi_{ij}$ the coupling tensor $f_{ij}$ can not be determined uniquely. In fact if $
f_{ij}$ is a solution then for any arbitrary unitary matrix $U$, $fU$ is also a solution. But this
freedom does not affects the physical observables. Therefore, we may take the
coupling tensor to be symmetric, i.e, $f_{il}f_{jl}=f_{ij}^2$, and inverse the relation (\ref{a12}) for a specified susceptibility tensor $\chi_{ij}$. Then, the corresponding coupling tensor up to a unitary freedom is as follows:
\begin{equation}\label{a13}
f_{ij}(\omega ) = \sqrt{\frac{{2\omega }}{\pi
}{\rm Im}{\chi}_{ij}(\omega)}.
\end{equation}
Now by substituting (\ref{a9}) into Eq. (\ref{a5}), the classical Langevin equation are obtained as
\begin{equation} \label{a14}
m{ \ddot q}_i(t) + \int_0^t dt'\,\dot {{\chi }}_{ij}(t -
t'){ \dot q}_j(t') +  \frac{\partial V({\rm q})}{\partial q_i}
={\xi}_i^{\rm N}(t),
\end{equation}
where $\boldsymbol{\xi}^{\rm N}(t)=-{\rm \dot{R}}^N(t)$. Here, the frictional force is a linear functional of the history of the velocity $\bf\dot q$ of the particle and the stochastic force $\boldsymbol{\xi}^{\rm N}(t)$ which is a result of the dissipation character of the medium and obeys the Gaussian statistics \cite{10,13}.
%
%
\section{Non-relativistic quantum dynamics}
The Lagrangian (\ref{a1}) can now be used to obtain the canonical conjugate variables correspond to the
dynamical variables ${\bf X}$ and ${\bf q}$ respectively as
\begin{eqnarray} \label{b1}
&&  Q_{i}(\omega,t)  = \frac{\delta L}{\delta \dot {X}_{ i }(\omega,t)
} = \dot {X}_i(\omega,t),\\
&&  p_i (t) = \frac{\delta L}{\delta \dot {q}_i } = m\dot {q}_i + {R}_i (t).\nonumber
\end{eqnarray}
The fields can be canonically quantized in a standard fashion by demanding equal-time commutation relations
among the variables and their conjugates
\begin{eqnarray} \label{b2}
 &&[q_i(t), p_j(t)] = \iota \hbar\, \delta _{ij},
\end{eqnarray}
\begin{equation}\label{b3}
[X_i(\omega,t), Q_j(\omega',t)] = \iota\, \hbar\, \delta _{ij}\, \delta (\omega  - \omega '),
\end{equation}
with all other equal-time commutators being zero. Using the
Lagrangian (\ref{a1}) and the expression for the canonical conjugate
variables in (\ref{b1}), we obtain the Hamiltonian of the total
system
\begin{eqnarray} \label{b4}
&&H= \frac{({\bf p} - {\bf R}(t))^2}{2m}+ V({\bf q})+\frac{1}{2} \int_0^\infty d\omega ({\bf Q}^2 (\omega,t)+
\omega ^2 {\bf X}^2(\omega,t)).
\end{eqnarray}
It is seen that this has the same form as a charged particle Hamiltonian that interacting with the electromagnetic field through a minimal coupling provided that the vector field ${\bf A}$ taking as the vector field ${\bf R}$. This result justify a minimal coupling scheme by making the substitution ${\bf p}\rightarrow({\bf p} - {\bf R}(t))$ to introduce the dissipation effect of the medium in Hamiltonian formalism~\cite{51}.\\
To facilitate calculations, let us introduce the following annihilation operator
\begin{eqnarray} \label{b19}
&& b_i (\omega ,t) = \sqrt {\frac{1}{2\hbar\omega}} [\omega X _{ i }(\omega,t)+ \imath Q _{i}(\omega,t)],
\end{eqnarray}
from equal-time commutation relations (\ref{b2})-(\ref{b3}), we find
\begin{eqnarray} \label{b20}
&& \left[ {b_i  (\omega ,t),b_{j}^ \dag (\omega ',t)} \right] = \delta _{ij}\, \delta (\omega  - \omega ').
\end{eqnarray}
Inverting the equation (\ref{b19}), we can write the canonical
conjugate variables  $ X_i (\omega,t)$ and $ Q_i (\omega,t)$ in terms of
the creation and annihilation
 operators  $b_i ^ \dag $ and $b_i $ as
\begin{eqnarray} \label{b21}
&&  { X} _{ i} (\omega,t) = \sqrt {\frac{\hbar }{{2\omega }}}
 {\left( {b_i (\omega ,t) + b_i ^ \dag (\omega
,t)}
\right)  } ,\nonumber \\
 &&  { Q} _{ i}
(\omega,t) = \iota \sqrt {\frac{{\hbar \omega }}{2}}  {\left( {b_i^
\dag  ( \omega ,t) - b_i (\omega ,t)} \right)} .
\end{eqnarray}
Using these relations, we can obtain the Hamiltonian of the total system in terms of the creation and
annihilation operators of the medium
\begin{eqnarray} \label{b22}
 H = \frac{{\left[ {{\bf p} -
{\bf R}(t) } \right]^2 }}{2m} + V({\bf q})  + H_{\rm m},
\end{eqnarray}
where
\begin{equation}\label{b23}
{{ R}_i}(t) =  {\int_0^\infty {d\omega }}\,\sqrt {\frac{\hbar }{{2\omega }}}f _{ij}(\omega )\left[ {b_j (\omega
,t) + b_j ^ \dag ( \omega ,t)} \right],
\end{equation}
and
\begin{eqnarray} \label{b24}
&& H_{\rm m} = :{\int {d\omega } } \,\hbar \omega \,\,b_i ^ \dag  (\omega ,t)\,b_i  (\omega ,t):
\end{eqnarray}
is the Hamiltonian of the medium in normal ordering form.
In the Heisenberg picture, by using commutation relations (\ref{b2}), (\ref{b3}) and the total Hamiltonian (\ref{b4}),
 the equations of motion for the canonical variables ${\bf X }$ and ${\bf Q}$ are obtained as
\begin{equation}\label{b5}
\dot {X}_i(\omega,t) = \frac{\iota }{\hbar }[H, X_i(\omega,t)] = Q_i(\omega,t),
\end{equation}
\begin{equation} \label{b6}
\dot {Q}_i(\omega,t)= \frac{\iota }{\hbar }[H, Q_i(\omega,t)] = - \omega ^2 X_i(\omega,t) + \dot {q}_j (t) f
_{ji}(\omega ).
\end{equation}
It can be easily shown that the combination of these equations leads to the same classical equation (\ref{a6}) with the formal solution (\ref{a8}).
In a similar way, using Heisenberg equation for the conjugate dynamical
variables ${\bf \dot q}(t)$ and ${\bf \dot p}(t)$, we find
\begin{eqnarray} \label{b7}
&& {\bf \dot q}(t) = \frac{\iota }{\hbar }\left[ {H,{\bf q}(t)}
\right] = \frac{\left( {{\bf p} - {\bf R}(t)} \right)}{m},
\end{eqnarray}
\begin{eqnarray} \label{b8}
&& {\bf \dot p}(t) = \frac{\iota }{\hbar }\left[ {H,{\bf p}(t)}
\right] =  - \boldsymbol{\nabla} V ({\bf q}).
\end{eqnarray}
Combination of these recent equations also lead to the same classical equation of motion (\ref{a5}). In addition, by substituting the solution (\ref{a8}) in the latter equation, the quantum analogous of the Langevin equation (\ref {a14}) are obtained as
\begin{equation} \label{b9}
m{ \ddot q}_i(t) + \int_0^t dt'\,\dot {{\chi }}_{ij}(t -
t'){ \dot q}_j(t') +  \frac{\partial V({\bf q})}{\partial q_i}
={\xi}_i^{\rm N}(t),
\end{equation}
where the susceptibility ${\chi}_{ij}$ has been already defined in Eq.~(\ref{a10}) and ${\xi}_i^{\rm N}(t)$ is a fluctuating force induced by the medium and in terms of the introduced operator~(\ref{b19}) are written as
\begin{equation} \label{b10}
{\xi}_{i}^{\rm N}(t) = \imath\int_0^\infty  {d\omega } \,\sqrt{\frac{\hbar\omega}{2}}\, f_{ij}(\omega)\left({
{b}_j(\omega,0)e^{-\imath \omega t} - {\rm h.c.}} \right).
\end{equation}
The Eq.~(\ref{b9}) is the quantum Langevin equation, wherein the explicit form of the noise is known. The
quantum Langevin equation can be considered as the basis of the macroscopic description of a quantum system
coupled to an environment or a heat bath. This equation contains a memory tensor $\dot{\chi}_{ij}(t)$ and a noise
or fluctuating force $\boldsymbol{\xi}^{\rm N}(t)$. If the medium is kept in thermal equilibrium at temperature $T$, by using Eq.~(\ref{b10}) the
force noise correlations~\cite{10} can be found as
\begin{eqnarray}\label{b11}
&&\zeta_{ij}(t-t')\equiv\frac{1}{2}\left\langle {\xi^{\rm N} _i (t)\xi^{\rm N} _j
(t') + \xi^{\rm N} _i (t')\xi^{\rm N} _j (t)} \right\rangle  =\nonumber\\
&& \int_0^\infty {d\omega } \frac{{\hbar \omega }}{\pi }{\mathop{
{\rm Im}}\nolimits} [\chi _{ij} (\omega )]\cos\omega (t - t')\coth
(\frac{{\hbar \omega }}{{2k_BT}}).
\end{eqnarray}
From this recent relation, the power spectrum of the noise force is obtained as
\begin{eqnarray}\label{b12}
\zeta_{ij}(\omega)\equiv\int_{ - \infty }^\infty
{dt}\, \zeta _{ij} (t)\cos \omega t = \hbar \omega \coth
\frac{{\hbar \omega }}{{2k_BT}}{\rm Im}[ \,\chi
_{ij} (\omega )],
\end{eqnarray}
which is a version of the quantum mechanical Fluctuation-dissipation theorem \cite{1,10,52}. It should be noted that the above equation
can be used to find a special susceptibility tensor in
order to have a predetermined correlation function, for example a
white noise. The limit $\hbar \rightarrow 0$ clearly gives a smooth transition to the classical Langevin equation, in the sense that in this
limit all commutators vanish and the equation of motion becomes an equation for c-numbers driven by a noise
term. Since $\mathop {\lim }\limits_{\hbar  \to 0} \,\hbar \omega \,\coth ({{\hbar \omega } \mathord{\left/
 {\vphantom {{\hbar \omega } {2k_BT}}} \right.
 \kern-\nulldelimiterspace} {2k_BT}}) = 2k_BT$, this
noise term possesses a flat spectrum \cite{10,13}.

In the case of a time-local friction proportional to the velocity that is usually called Ohmic, i.e,
$\chi_{ij}(t)=\gamma\delta_{ij}\theta(t) $, we find that $f_{ij}(\omega)=\sqrt{\frac{2\gamma}{\pi}}\delta_{ij}$
and the force noise correlations (\ref{b11}) are simplified as
\begin{equation}\label{b13}
\zeta_{ij}(t-t')  = \gamma
k_BT\frac{d}{{dt}}{\coth \big[\frac{{\pi k_BT(t - t')}}{\hbar }\big]}\delta_{ij},
\end{equation}
which is consistent with the results have been reported in
\cite{10,13,52}. In the remainder of the paper we consider four
examples to show the applicability of this scheme.
%
%
\subsection{Free particle}
The simplest model for a dissipative Brownian particle is described by a free
particle. We assume here, the particle with mass $m$ is moving in an anisotropic dissipative medium. This model
can be applied to the problem of Cherenkov radiation in presence of a polarizable medium \cite{53,70}. We set $V({\bf q})=0$ in the equation~(\ref{b9}) and find
\begin{equation} \label{1}
m{\ddot q}_i(t) + \int_0^t dt'\dot\chi_{ij}(t - t'){ \dot q}_j(t') = \xi_i^{\rm N}(t).
\end{equation}
For any function $q(t)$ the forward- and the backward-Laplace-transform of $q(t)$ are respectively defined by
\begin{eqnarray}
&&\tilde q^f (s) = \int_0^\infty  {dt}\, e^{ - st} q(t),\nonumber\\
&&\tilde q^{\,b} (s) = \int_0^\infty  {dt}\, e^{ - st} q(-t).\nonumber
\end{eqnarray}
Depending on the initial conditions which are usually set at $t=0$, there is no need to use backward Laplace
transform but here we do not restrict ourselves and the reader can choose the plus sign in what follows to
recover this situation. Using these definitions, we take the Laplace transform of the both sides of the equation
(\ref{1}) as
\begin{eqnarray}
\Lambda _{ij}(s) \tilde q_j^{f,b}(s) = \frac{1}{s}\Lambda _{ij} q_j (0) \pm m\dot{q}_i (0) + \imath\int_0^\infty
{d\omega } \sqrt{\frac{{\hbar }}{2 \omega}} f_{ij} (\omega )\left( {\frac{b_j (\omega )}{{s \pm \iota \omega }} - h.c.} \right), \nonumber\\
\end{eqnarray}
where $\Lambda _{ij}(s) = \left( {ms^2 \delta _{ij}  + s^2 \tilde \chi _{ij} (s)} \right)$ and the upper(lower)
sign corresponds to $q_j^{f}(s)\,(\,q_j^{b}(s)\,)$. After some simple but elaborate calculations
one finds
\begin{equation}\label{2}
q_i (t) = q_i (0)\pm \eta_{ij}(t)p_j (0)+\int_0^\infty {d\omega }
\sqrt {\frac{{\hbar  }}{2\omega}} ( {Z^{\pm}_{ij}(\omega , \pm
t)\,b_j (\omega ,0)} + h.c.),
\end{equation}
where now the upper(lower) sign corresponds to $t>0$ ($t<0$) and the functions $\eta_{ij}(t)$ and $Z^{\pm }_{ij}(\omega ,\pm t)$
are defined by
\begin{eqnarray}\label{3}
\eta_{ij}(t)&=&L^{ - 1} \left[ {\Lambda_{ij}^{-1}(s)} \right],\\
Z^{\pm }_{ij}(\omega , \pm t)& =& L^{ - 1} \left[
{\Lambda_{ik}^{-1}(s)\frac{\mp s}{s\pm\imath\omega}}
\right]f_{kj}(\omega),
\end{eqnarray}
in which, $L^{-1}[f(s)]$ denotes the inverse Laplace transform of
function $f(s)$ and $\Lambda^{-1}(s)$ is the inverse of the matrix
$\Lambda(s)$.\par
Diffusion of a Brownian particle is
characterized by the long-time behavior of the mean-square
displacement, therefore we compute the mean square distance
traveled by the free particle in a time interval starting at the
time $t'$ and ending at the time $t$
\begin{eqnarray}\label{m1}
&&\left\langle {[{\bf q}(t) - {\bf q}(t')]^2 } \right\rangle  = \sum_{i=1}^{3}(\eta _{ij} (t) - \eta _{ik}
(t'))\left\langle {p_j (0)p_k (0)}
\right\rangle \nonumber \\
&&+ \int_0^\infty  {d\omega } \frac{{\hbar \omega }}{2}\coth
\frac{{\hbar \omega }}{{2KT}}\{ [Z^\pm_{ij} (\omega ,t) - Z^\pm_{ij}
(\omega ,t')][Z^{*\pm}_{ji}  (\omega ,t) - Z^{*\pm}_{ji}  (\omega ,t')]  \},\nonumber \\
\end{eqnarray}
which in the case of the Ohmic damping is reduced to
\begin{eqnarray}\label{m1}
&&\left\langle {[{\bf q}(t) - {\bf q}(t')]^2 } \right\rangle  = \int_0^\infty  {d\omega } \frac{{\hbar \gamma
}}{{\pi \omega }}\frac{{\coth ({{\hbar \omega } \mathord{\left/ {\vphantom {{\hbar \omega } {2KT}}} \right.
\kern-\nulldelimiterspace} {2KT}})}}{{\gamma ^2  + m^2 \omega ^2 }}\nonumber \\
&& (e^{ - \iota \omega t}  - e^{ - \iota \omega t'}  - e^{ -
\frac{\gamma }{m}t}  + e^{ - \frac{\gamma }{m}t'} )(e^{\iota
\omega t}  - e^{\iota \omega t'}  - e^{ - \frac{\gamma }{m}t}  +
e^{ - \frac{\gamma }{m}t'}).
\end{eqnarray}
This integral contains transient terms, which appears to allow the particle to absorb a large amount of energy
from medium, however after this transient, for sufficiently large times, diffusion of the particle is given by a
formula which is known in classical stochastic theory. In large-time-limit we find
\begin{eqnarray}\label{m2}
&&\left\langle {[{\bf q}(t) - {\bf q}(t')]^2 } \right\rangle  \approx \frac{{\hbar \gamma }}{\pi }\int_0^\infty
{d\omega } \frac{{\omega \coth ({{\hbar \omega } \mathord{\left/ {\vphantom {{\hbar \omega } {2KT}}} \right.
\kern-\nulldelimiterspace} {2KT}})}}{{\gamma ^2  + m^2 \omega ^2 }} \frac{{4\sin ^2 \frac{\omega }{2}(t -
t')}}{{\omega ^2 }},
\end{eqnarray}
from which the following asymptotic formula can be found \cite{10,13,40}.
\begin{equation}\label{m3}
\left\langle {[{\bf q}(t) - {\bf q}(t')]^2 } \right\rangle \approx \frac{{4KT}}{\gamma }(t - t').
\end{equation}
%
%
\subsection{Harmonic oscillator}
The problem of a damped harmonic quantum mechanical oscillator has
been studied intensively  because of its universal
relevance~\cite{54}-\cite{58}. Indeed, it describes a quantum
electromagnetic field propagating in a linear dielectric medium
\cite{59}, a particle interacting with a quantum field in dipole
approximation \cite{60}. In addition to these quantum optical
applications, this problem is used in nuclear physics \cite{61}
and quantum chemistry \cite{1,62}. For a harmonic oscillator with
mass $m$ and frequency $\omega_0$, we set $V({\bf
q})=\frac{1}{2}m\omega_0^2{\bf q}^2$ in the equation~(\ref{b9})
and find
\begin{equation} \label{4}
m{\ddot q}_i(t) + \int_0^t dt' \dot\chi_{ij}(t - t'){ \dot q_j}(t')+m\omega_0^2q_i(t) = {\xi_i}^{\rm N}(t),
\end{equation}
with the following solution
\begin{eqnarray}\label{5}
q_i (t) = \alpha_{ij}(t)q _j(0)\pm \eta'_{ij}(t)p_j
(0)+\int_0^\infty {d\omega }\sqrt {\frac{{\hbar
}}{2\omega}}({Z'^{\pm }_{ij}(\omega ,\pm t)\,b_j(\omega ,0)+ h.c.}),\nonumber \\
\end{eqnarray}
in which the upper (lower) sign again corresponds to $t>0$ ($t<0$). The
functions $\alpha_{ij}(t)$, $\eta'_{ij}(t)$ and $Z'^{\pm }_{ij}(\omega
,\pm t)$ are defined by
\begin{eqnarray}\label{6}
\alpha_{ij}(t)&=&L^{-1}[{s{\Lambda'}_{ik}^{-1}(s)(m\delta_{kj}+\chi_{kj}(s))}
],\\
\eta'_{ij}(t)&=&L^{ - 1}[{{\Lambda'}_{ij}^{-1}(s)}],\\
Z'^{\pm }_{ij}(\omega , \pm t)& =& L^{ - 1}[
{{\Lambda'}_{ik}^{-1}(s)\frac{\mp
s}{s\pm\imath\omega}}]f_{kj}(\omega),
\end{eqnarray}
with
\begin{equation}
{\Lambda'}_{ij}(s)=ms^2\delta_{ij}+s^2\chi_{ij}(s)+m\omega_0^2.
\end{equation}
During the last decade huge advances in laser cooling and trapping
experimental techniques have made it possible to confine
harmonically a single ion and cool it down to very low
temperatures where purely quantum manifestations begin to play an
important role \cite{63}. The reference \cite{64} reports recently
experimental results on how to couple a properly engineered
reservoir with a quantum oscillator. These experiments aim at
measuring the decoherence of a quantum superposition of coherent
states and Fock states due to the presence of the reservoir
\cite{58}. Therefore, in order to survey the theory bases of this kind of issue, we calculate the transition probabilities of an initially excited harmonic oscillator embedded in an
anisotropic dissipative medium. For this purpose we write the
Hamiltonian (\ref{b22}) as
\begin{equation}\label{6.9}
H = H_0  + H' ,
\end{equation}
where
\begin{eqnarray}\label{7}
H_0  &=& \frac{{{\bf p}^2 }}{{2m}}+ \frac{1}{2}m\omega_0^2{\bf q}^2+ H_{\rm m} \nonumber \\
&=&\hbar\omega_0(a^\dag_ja_j+\frac{3}{2})+ H_{\rm m},
\end{eqnarray}
with
\begin{equation}\label{7/1}
H'= - \frac{{{\bf p} \cdot {\bf R}}}{m} + \frac{{{\bf R}^2
}}{{2m}}.
\end{equation}
The term $\frac{{{\bf R}^2}}{{2m}}$ can be ignored for a sufficiently weak coupling and the
dominant term in this case is $\frac{{{\bf p} \cdot {\bf R}}}{m}$.
Using the rotating-wave approximation \cite{63,64}, $H'$ in the interaction picture can be
written as
\begin{eqnarray}\label{8}
H'_{\rm I}=\iota \sqrt {\frac{{\hbar \omega _0 }}{{2m}}} \int_0^\infty {d\omega } f_{ij}(\omega )[{a_i (0)\,b_j^\dag
(\omega ,0)e^{-\imath(\omega_0-\omega)t} - a_i^\dag (0)\,b_j (\omega
,0)e^{\imath(\omega_0-\omega)t}}].\nonumber\\
\end{eqnarray}
In order to obtain the transition-probabilities, we find the density operator to using the time-dependent
perturbation theory. The time-evolution of the density-operator in the interaction-picture can be obtained from
\cite{65}-\cite{68}
\begin{equation}\label{9}
\rho _{\rm I} (t) = U_{\rm I}^\dag (t)\rho _{\rm I} (0)U_{\rm I} (t),
\end{equation}
where $ U_{\rm I} (t)$, up to the first order time-dependent perturbation is
\begin{eqnarray}\label{10}
U_{\rm I} (t) &=& 1 - \frac{\iota }{\hbar }\int_0^\infty  {dt'} H_{\rm I} (t')\nonumber\\
&=&1+\sqrt {\frac{{ \omega _0 }}{{2m\hbar}}}\int_0^\infty {d\omega }
f_{ij}(\omega )\left( {a_i (0)b_j^\dag (\omega
,0)e^{-\imath(\frac{\omega_0-\omega}{2})t}} \right.\nonumber\\
&&\hspace{3cm} -\left. {a_i^\dag (0)b_j (\omega ,0)e^{\imath(\frac{\omega_0-\omega}{2})t}} \right)
\frac{sin(\frac{\omega_0-\omega}{2})t}{(\frac{\omega_0-\omega}{2})}.
\end{eqnarray}
Let the initial density-operator be $\rho_{\rm I}(0)=\left| {n_j } \right\rangle_{\omega_0\,\omega_0}\left\langle {n_j
} \right| \otimes \left| 0 \right\rangle _{m\,m} \left\langle 0 \right|$ where $\left| 0 \right\rangle _m $ is the
vacuum state of the dissipative medium and $\left| n_j \right\rangle _{\omega_0} $ an excited state of the
harmonic oscillator for the mode $j,\,(j=1,2,3)$. Then, by substituting $U_{\rm I} (t)$ from (\ref{10}) in (\ref{9})
and tracing out the medium degrees of freedom, the transition probabilities $ \left| n_j \right\rangle
_{\omega_0} \to \left| {n_j \pm 1} \right\rangle _{\omega_0} $ are obtained as
\begin{eqnarray}\label{11}
\Gamma _{n_j \to n_j + 1} & =& {\rm Tr_s} \left[ {\left( {\left| {n_j +
1} \right\rangle _{\omega _0\,\omega_0 } \left\langle {n_j + 1}
\right|}
\right){\rm Tr_m} [\rho_{\rm I} (t)]} \right] = 0 ,\\
\Gamma _{n_j \to n_j - 1}  &=& {\rm Tr_s} \left[ {\left( {\left| {n_j -
1} \right\rangle _{\omega _0\,\omega_0 } \left\langle {n_j - 1}
\right|}
\right){\rm Tr_m} [\rho_{\rm I} (t)]} \right] \nonumber\\
&=& \frac{{\omega _0 n_j}}{{m\pi\hbar }}\int_0^\infty  {d\omega }\,\omega\,{\rm Im} [\chi_{jj}(\omega )]
\frac{\,{\sin ^2 \frac{{(\omega  - \omega _0
)}}{2}t}}{{\left( {\frac{{\omega  - \omega _0 }}{2}} \right)^2
}} ,
\end{eqnarray}
in which ${\rm Tr_m}\,({\rm Tr_m})$ denotes that the trace is taken over the degrees of freedom of the medium(harmonic oscillator). In the large-time-limit, $\frac{{\sin ^2 \frac{{(\omega - \omega _0 )}}{2}t}}{{\left( {\frac{{\omega  - \omega _0
}}{2}} \right)^2 }}\approx 2\pi t\delta(\omega-\omega_0)$, therefore
\begin{equation}\label{12}
\Gamma _{n_j \to n_j - 1}  = \frac{{2\omega _0^2 n_j t }}{{m\hbar
}}{\rm Im} [\chi_{jj}(\omega_0 )].
\end{equation}
In a similar way, when the medium has a Maxwell-Boltzman distribution, i.e, $ \rho_{\rm m}= \frac{{e^{\frac{{ - H_{\rm m}
}}{{k_{\rm B}T}}} }}{{
{\rm Tr_m} [e^{\frac{{ - H_{\rm m} }}{{k_{\rm B}T}}} ]}}$ and $\rho_{\rm I}(0)=\left| {n_j }
\right\rangle_{\omega_0\,\omega_0}\left\langle {n_j } \right| \otimes \rho_m$, then the
transition-probabilities $ \left| n_j \right\rangle _{\omega_0} \to \left| {n_j \pm 1} \right\rangle
_{\omega_0}$, in the large-time-limit, can be obtained as
\begin{eqnarray}\label{13}
\Gamma _{n_j \to n_j + 1}  = \frac{{2 \omega _0^2 (n_j + 1)\overline{n}(\omega,T)t}}{{m\hbar }}
{\rm Im} [\chi_{jj}(\omega_0 )],
\end{eqnarray}
\begin{equation}\label{13.1}
\Gamma _{n_j \to n_j - 1}  = \frac{{2 \omega _0^2 n_j\,(\overline{n}(\omega,T)+1)
t }}{{m\hbar }}{\rm Im} [\chi_{jj}(\omega_0 )],
\end{equation}
where $\overline{n}(\omega,T)=[\exp({\frac{{\hbar\omega _0
}}{{k_{\rm B}T}}})  - 1]^{-1}$ is the mean number of thermal photons at frequency $\omega$ and
temperature $T$. It is clear from above equations, when the medium is held at zero temperature then $\Gamma _{n_j \to n_j + 1} =0$, and as expected the energy flows only from the system to the medium until the system falls in
its ground state and remains in this state forever.
%
\subsection{Spontaneous emission of an excited two-level atom}
In this section we calculate the decay constant and level shift for an initially excited atom embedded in an
anisotropic medium. Theoretical treatments often simplify matters even more by assuming that the atoms only have
two levels and is localized, which experimentally can only be achieved by rather complicated optical pumping
techniques~\cite{67}. Indeed it describes systems composed of atoms interacting with a few modes of the electromagnetic
field which has been studied both theoretically and experimentally \cite{13},\cite{66}-\cite{68}. For this
purpose, let us consider a one-electron atom and write the Hamiltonian (\ref{b22}) as
\begin{equation}\label{c1.1}
H = H_0  + H' ,
\end{equation}
where
\begin{equation}
H_0  = H_{Atom}  + H_{\rm m} , \\
\end{equation}
and
\begin{eqnarray} \label{c1}
H' =   {\bf q} \cdot {\bf \dot R}(t),
\end{eqnarray}
with
\begin{equation}\label{c2}
H_{Atom}  = \frac{{{\bf p}^2 }}{{2m}} + V({\bf q}).
\end{equation}
For a two-level atom with upper state $\left| 2 \right\rangle $,
lower state $\left| 1 \right\rangle $  and transition frequency
$\omega_0$, the Hamiltonian (\ref{c1.1}) can be written as
\begin{equation}\label{c3}
H = \hbar \omega _0 \sigma ^\dag  \sigma    + ({\bf q}_{\,12}
\,\sigma  + {\bf q}_{\,12}^ *  \,\sigma ^\dag  )\cdot{\bf\dot{R}
}(t)+ H_{\rm m}
\end{equation}
where $\sigma  = \left| 1 \right\rangle \left\langle 2 \right| $
and $\sigma^\dag  = \left| 2 \right\rangle \left\langle 1 \right|
$  are Pauli operators of the two-level atom and ${\bf q}_{\,12} =
\left\langle 1 \right|{\bf q}\left| 2 \right\rangle$ is its
transition dipole momentum. To study the spontaneous decay of an
initially excited atom we may look for the wave function of total
system as follows
\begin{equation}\label{c4}
\left| {\psi (t)} \right\rangle  = c(t)\left| 2 \right\rangle \left| 0 \right\rangle _m  + \sum\limits_{i = 1}^3
{\int_0^\infty {d\omega } } M_i (\omega,t )\left| 1 \right\rangle \left| 1_i(\omega ) \right\rangle _m
\end{equation}
where $\left| 0 \right\rangle _m$ and $\left| 1_i(\omega )\right\rangle _m$ are the vacuum state and the excited state of the medium with a single photon with frequency $\omega$ and polarization $i$, respectively. The coefficients $c(t)$ and $M_i (\omega,t
)$ are to be specified by the Schrodinger equation
\begin{equation}\label{c5}
\imath\hbar \frac{{\partial \left| {\psi (t)} \right\rangle
}}{{\partial t}} = H\left| {\psi (t)} \right\rangle,
\end{equation}
with the initial conditions $c(0)=1$, $M_i (\omega,0 )=0$. According to the relations are shown in appendix, the spontaneous decay of an initially excited atom are given by
\begin{eqnarray}\label{c6}
c(t)=e^{-\Gamma t}e^{-\imath(\Delta+\omega_0)t},
\end{eqnarray}
where
\begin{equation}\label{c7}
\Gamma = \frac{\omega_0^2}{\hbar}\left[{{\rm q}_{12,i}^*\,
{\rm Im}[{ \chi}_{ij}(\omega_0 )]\,{\rm q}_{12,j}} \right],
\end{equation}
and
\begin{equation}\label{c8}
\Delta = P\int_0^\infty {d\omega } \frac{{q}_{12,i}^*\,
{\rm Im}[{ \chi}_{ij}(\omega )]\,{\rm q}_{12,j}}{{\pi\hbar(\omega_0 - \omega )}},
\end{equation}
are the decay constant and the level shift due to the presence of the dissipative medium, respectively. The
symbol $P$ denotes the Cauchy principal value.

In the case of an isotropic susceptibility such as the lorenz model, i.e. $\chi_{ij} (t) = \beta e^{\frac{{ - \gamma t}}{2}} \frac{{\sin \nu t}}{\nu }\delta_{ij}$ which $\beta$ and $\nu$ are a positive constants and $\gamma$ is a damping coefficient, we find that $f^2 (\omega ) =  {\frac{{\beta \omega }}{{2\pi \nu }}} \left[{{\frac{\gamma }{{\frac{{\gamma ^2 }}{4} + (\nu  + \omega )^2 }} - \frac{\gamma }{{\frac{{\gamma ^2 }}{4} + (\nu  - \omega )^2 }}}}\right ]$. This recent coupling function in nondissipative limit, $\gamma\rightarrow 0$, are reduced to $f^2 (\omega ) = \frac{{\beta \omega }}{\nu }\delta (\nu  - \omega )$. Accordingly, by inserting it into Eqs. (\ref{c7}) and~(\ref{c8}), we obtain
\begin{eqnarray}\label{c12}
&& \Gamma  = \frac{{\pi \beta\omega _0^2 \left| {{\bf q}_{12} }
\right|^2
}}{{2\hbar\nu }}\delta (\nu  - \omega _0 ),\\
&& \Delta  = \frac{{\beta \nu \left| {{\bf q}_{12} } \right|^2
}}{{2\hbar(\omega _0  - \nu )}},
\end{eqnarray}
which is consistent with the results have been reported in~\cite{66}-\cite{68}.
%
%
\subsection{Radiation reaction}
A charge particle in accelerated motion radiates electromagnetic waves, and
as a result it experiences a friction-like force, i.e, radiation
reaction. Consider a charged particle embedded in an
anisotropic-dissipative-medium interacting with its own field and
also the quantum vacuum field. The total Lagrangian that describe this system is a generalized version of the Lagrangian~(\ref{a1}) by adding the Lagrangian electromagnetic field together with its interaction with the charged particle. Therefore, the total Lagrangian in Coulomb-gauge can be written as
\begin{eqnarray}\label{f4}
&&L=\frac{1}{2}m\,{\bf \dot q}^2(t)+\frac{1}{2}\int_0^\infty
{d\omega [{\bf \dot X}^2(\omega,t)- \omega ^2 {\bf X}^2(\omega,t) ]}\nonumber\\
&&+\frac{1}{2}\int {d^3{\bf r }[\varepsilon _0 {\bf \dot{A}}^2  -
\frac{{(\nabla\times{\bf A})^2 }}{{\mu _0 }}]}-\frac{1}{2}\int
d^3{\bf r }\rho({\bf r})\phi({\bf r},t)+{\bf \dot{q}}\cdot{\bf R}(t).
\end{eqnarray}
By using Lagrangian (\ref{f4}), the corresponding canonical conjugate variables can be found
\begin{eqnarray} \label{f5}
-\varepsilon_0E_i^\bot({\bf r},t)  = \frac{\delta L}{\delta \dot {A}_{ i }({\bf
r},t) } =\varepsilon_0 \dot {A}_i({\bf r},t),
\end{eqnarray}
\begin{equation}\label{f6}
p_i (t) = \frac{\delta L}{\delta \dot {q}_i } = m\dot {q}_i + {R}_i
(t)+e A_i({\bf r},t).
\end{equation}
The electromagnetic field can be canonically quantized by imposing the following equal-time
commutation relation
\begin{eqnarray} \label{f7}
\left[ { {A} _i ({\bf r},t),-\varepsilon_0E_j^\bot ({\bf r'},t)} \right] =
\iota \hbar \delta _{i j} \delta^\bot ({\bf r} - {\bf r'}).
\end{eqnarray}
where $\delta^\bot ({\bf r} - {\bf r'})$ is the transverse delta function. To facilitate the calculations, let us introduce new annihilation-operators
\begin{eqnarray} \label{f8}
a_\lambda  ({\bf k},t) = \sqrt {\frac{c|{\bf k}|\varepsilon _0}{{2\hbar
 }}} \left( { c|{\bf k}|
\underline A _\lambda  ({\bf k},t) - \iota \underline E_\lambda ^
\bot  ({\bf k},t)} \right),
\end{eqnarray}
where $\underline { A}_\lambda ({\bf k},t)$ and $\underline { E}_\lambda^\bot ({\bf k},t)$ are the cartesian components of the spatial Fourier transformation of $ {\bf A} ({\bf r},t)$ and $ {\bf E}^\bot ({\bf r},t)$. From equal-time commutation relation~(\ref{f7}), the following equal-time commutation relations are obtained
\begin{eqnarray} \label{f9}
\left[ {a_\lambda  ({\bf k},t),a_{\lambda '}^ \dag  ({\bf k'},t)}
\right] = \delta _{\lambda \lambda '} \delta ({\bf k} - {\bf k'}).
\end{eqnarray}
Inverting Eq.~(\ref{f8}) and taking the inverse Fourier transform, we obtain the electromagnetic field operators in
terms of the creation and annihilation operators in the real-space as
\begin{eqnarray} \label{f10}
&& {\bf A}({\bf r},t) = \sum\limits_{\lambda  = 1}^2 {\int {d^3 {\bf
k}} \sqrt {\frac{\hbar }{{2(2\pi )^3 \varepsilon _0 c|{\bf k}| }}} }
\left( {a_\lambda  ({\bf k},t)e^{\iota {\bf k} \cdot {\bf r}} +
a_\lambda ^ \dag  ({\bf k},t)e^{ - \iota {\bf k}
\cdot {\bf r}} } \right){\bf e}_\lambda  ({\bf k}),\nonumber \\
&& {\bf E}^ \bot  ({\bf r},t) =  - \iota
\sum\limits_{\lambda  = 1}^2 {\int {d^3 {\bf k}} \sqrt {\frac{{\hbar
c|{\bf k}| }}{{2(2\pi )^3 \varepsilon _0 }}} } \left( {a_\lambda ^ \dag
({\bf k},t)e^{ - \iota {\bf k} \cdot {\bf r}}  - a_\lambda  ({\bf
k},t)e^{\iota {\bf k} \cdot {\bf r}} } \right){\bf e}_\lambda  ({\bf
k}),\nonumber
\end{eqnarray}
where ${\bf e}_\lambda  ({\bf k}),\, (\lambda=1,2)$ are orthonormal polarization vectors. Using Lagrangian (\ref{f4})
and the expressions for the canonical conjugate variables in (\ref{b2}) and (\ref{b20}), we find Hamiltonian of the
total system as
\begin{eqnarray} \label{f11}
 H = \frac{{\left( {{\bf p} -
{\bf R}(t) - e{\bf A}({\bf q},t)} \right)^2 }}{2m} + \frac{1}{2}\int
d^3{\bf r }\rho({\bf r})\phi({\bf r},t)+ H_{\rm F}  + H_{\rm m},
\end{eqnarray}
where
\begin{eqnarray} \label{f12}
&&H_{\rm F}  = :\sum\limits_{\lambda  = 1}^2 {\int {d^3 {\bf k}} \,\,\hbar
c |{\bf k}| \,a_\lambda ^ \dag  ({\bf k},t)a_\lambda ({\bf k},t):}
\end{eqnarray}
is the Hamiltonian of the electromagnetic field in the normal ordering form and $H_m$ is the Hamiltonian of the
heat bath which is already defined in Eq.~(\ref{b24}).

In the Heisenberg picture, the equations of motion for canonical variables $ {\bf X} _{\omega }$ and ${\bf  Q} _{\omega }$
are the same equations (\ref{b5}) and (\ref{b6}) with the formal solution (\ref{b8}) and likewise the ${\bf R}$ field
is defined by Eq. (\ref{b9}). In a similar way, by combining the Heisenberg equations for the conjugate dynamical variables ${\bf \dot q}(t)$ and ${\bf \dot p}(t)$ we find
\begin{eqnarray} \label{f13}
&&m{ \ddot q}_i(t) +  \int_0^t dt' \dot {{\chi }_{ij}}(t -
t'){ \dot q}_j(t') +e\frac{\partial\phi({\bf q})}{\partial q_i} =e{ E}_i ({\bf q},t)\nonumber\\&& + e\, \epsilon_{ijk}\,{ \dot q}_j(t)  { B}_k({\bf q},t)+ {{ \xi} }_i^{\rm N}(t),
\end{eqnarray}
where the $\epsilon_{ijk}$ are the components of the Levi-Civita pseudotensor and the susceptibility tensor ${\chi }_{ij} $ and the noise operator ${ \xi}_i^{\rm N}(t)$ are the same previously equations where defined by Eqs.~(\ref{a10}) and~(\ref{b10}). Consider now a single electron with binding potential
energy $V({\bf r})=e\phi({\bf r})$. Suppose that the distances over which the bound electron can move in this
potential are small compared with the wavelength of any field with the electron undergoes a significant
interaction. Therefore, it is convenient to make the electric dipole approximation in which spatial variation of ${\bf
A}$ is ignored. Using this note, the Heisenberg equation for the operator $a_\lambda({\bf k},t)$ is found from the Hamiltonian (\ref{f11}) to be
\begin{equation}\label{f14}
\dot a_\lambda  ({\bf k},t) =  - \iota \omega _{\bf k} a_\lambda
({\bf k},t) + \iota e\frac{{\,\dot {\bf q}(t) \cdot {\bf e}_\lambda
({\bf k})}}{{\sqrt {2(2\pi )^3 \varepsilon _0 \hbar \omega _{\bf k} } }},
\end{equation}
with the following formal solution
\begin{equation}\label{f15}
a_\lambda  ({\bf k},t) = e^{ - \iota \omega _{\bf k} t} a_\lambda
({\bf k},0) + \frac{{\iota e\,}}{{\sqrt {2(2\pi )^3 \varepsilon _0
\hbar \omega _{\bf k} } }}\int_0^t {dt'} \,e^{ - \iota \omega _{\bf
k} (t - t')} {\bf e}_\lambda ({\bf k}) \cdot \dot {\bf q}(t'),
\end{equation}
where $\omega_{\bf k}=ck$. Now, by inserting ${a}_\lambda  ({\bf k},t)$ into the right-hand side of
(\ref{f13}), we obtain
\begin{eqnarray} \label{f16}
m{\ddot q}_i(t) +  \int_0^t dt' \dot {{\chi }_{ij}}(t -
t'){ \dot q}_j(t') +e\frac{\partial\phi({\bf q})}{\partial q_i} = e{\rm E}_{0,i}^\bot
(t) +e{\rm E}_{RR,i}^\bot(t)+ {\xi}_i^{\rm N}(t),
\end{eqnarray}
where ${\rm E}_{0,i}^\bot$ and ${\rm E}_{RR,i}^\bot$ are, respectively, the components of the vacuum field and the radiation reaction field and define as follow:
\begin{equation}\label{f17}
{\bf E}_{0}^ \bot  (t) = \iota \sqrt {\frac{\hbar }{{2(2\pi )^3
\varepsilon _0 }}} \sum\limits_{\lambda  = 1}^2 {\int_0^\infty
{d^3{\bf k}} } \sqrt {\omega _{\bf k} } \left( {a_\lambda  ({\bf
k},0)e^{ - \iota \omega_{\bf k} t}  - a_\lambda ^\dag ({\bf
k},0)e^{\iota \omega_{\bf k} t} } \right){\bf e}_\lambda  ({\bf k}),
\end{equation}
\begin{eqnarray}\label{f18}
{\bf E}_{RR}^ \bot  (t)& = & - \frac{e}{{(2\pi )^3 \varepsilon _0
}}\sum_{\lambda=1}^2\int_0^t {dt'} \cos \omega _{\bf k} (t -
t')\int {d^3 {\bf k}} \,{\bf e}_\lambda  ({\bf{ k}}) \cdot
{\bf{\dot q}}(t')\nonumber\\
&=&\tau{\bf q}^{{\hspace{-0.25cm}}\cdots} (t) -\frac{\delta m}{e}\ddot{{\bf
q}}(t).
\end{eqnarray}
Here, $\tau=\frac{e}{{6\pi c^3 }}$ and $\delta m=\frac{e^2}{3\pi^2\epsilon_0 c^3}\int _0^\infty d\omega$ \cite{68}. The mass $m$ in Eq. (\ref{f16}) is the mass
of a bare particle which does not interact with electromagnetic field. It is fictitious, since the interaction
cannot be turned off. The experimental mass of the particle must include the present interaction with
field. Therefore, $\delta m$ is effectively a contribution to the mass and arises from the action of its own
field, namely, from the radiation reaction.
%
%
\section{Dissipative Dirac field}
The description of particles used in the preceding sections is
valid only when the particles are moving at velocities small
compared to the velocity of light. In this section, we generalize the preceding
formalism to describe relativistic particles embedded in an anisotropic-dissipative-medium. For this purpose, we use the following Lagrangian for the Dirac field under influence of a potential $V$ and its interaction with
an external electromagnetic field and the dissipative medium
\cite {70,68}
\begin{equation}\label{d0}
L=L_{\rm m}+L_{\rm s}+L_{\rm int},
\end{equation}
where
\begin{eqnarray} \label{d1}
L_{\rm s} &=& \frac{{\iota \hbar c}}{2}\int {d^3 {\bf x}} \left[
{\sum\limits_{\mu = 0}^3 {\left( {\bar \psi ({\bf x},t)\gamma ^\mu
\frac{{\partial \psi ({\bf x},t)}}{{\partial x^\mu  }} -
\frac{{\partial \bar \psi ({\bf x},t)}}{{\partial x^\mu  }}\gamma
^\mu  \psi ({\bf x},t)} \right)}} \right.
\nonumber \\
&&\left. {-\left( {mc^2+V({\bf x})  } \right)\bar\psi \psi } \right],
\end{eqnarray}
and
\begin{eqnarray}\label{d1/1}
L_{\rm int}& =& e\int {d^3 {\bf x}} \left[ {
{(c\bar \psi ({\bf x},t)\gamma ^j \psi ({\bf x},t){\rm A}^j ({\bf
x},t))}  - \bar \psi ({\bf x},t)\gamma ^0 \psi ({\bf x},t)\varphi
({\bf x},t)} \right]\nonumber\\
&&+c \int_0^\infty d\omega \int {d^3 {\bf x}}f_{ij}(\omega)\bar
\psi({\bf x},t)\gamma^i \psi({\bf x},t)X_j(\omega).
\end{eqnarray}
Here, $\gamma^\mu ,\, (\mu=0,...,3)$, are the Dirac matrices with $\gamma
^0 \equiv \beta $, $\gamma ^j \equiv \beta \alpha _j $ and $\bar \psi  \equiv \psi
^\dag \beta$. In a standard representation we have
\begin{equation}\label{d2}
\beta  = \left( {\begin{array}{*{20}c}
   I & 0  \\
   0 & { - I}  \\
\end{array}} \right),\,\,\,\,\,\alpha _j  = \left( {\begin{array}{*{20}c}
   0 & {\sigma _j }  \\
   {\sigma _j } & 0  \\
\end{array}} \right),
\end{equation}
where $\sigma _j,\, (j=1,2,3)$, are Pauli spin matrices and $I$ is the unit matrix. The Lagrangian~(\ref{d0}) is the relativistic generalization of the nonrelativistic Lagrangian~(\ref{f4}) in the Coulomb gauge. The part of the Lagrangian attributable to the transverse field, involving the energy density of electromagnetic field, has the same form as Eq.~(\ref{f4}), as does the part associated with the medium and the part corresponded to the instantaneous Coulomb interactions among the charged particles. (In the nonrelativistic Lagrangian the Coulomb interactions are written
explicitly in the form appropriate for classical point particles.) Here, the medium Lagrangian~(\ref{a3}) remain without any change, since the medium is stationary on the one hand. On the other hand, we often deal with a non-relativistic medium, although the relativistic version of the Lagrangian~(\ref{a3}) can be simply considered~\cite{71}. The main difference between the relativistic and nonrelativistic Lagrangian lied in the treatment of the particle kinetic term~(\ref{d1}) and it's interaction terms~(\ref{d1/1}). We proceed along the lines of the preceding section and define the canonical conjugate variable of the Dirac particle as
\begin{equation}\label{d2/1}
\frac{\partial L}{\partial \dot{\psi}}=\frac{\imath\hbar}{2}\psi^\dag.
\end{equation}
The Dirac field is quantized by imposing anti-commutation relations among the field components
\begin{eqnarray}\label{d2/2}
\{\psi_\alpha({\bf x},t),\psi^\dag_\beta({\bf x'},t)\}&=& \delta_{\alpha\beta}\delta({\bf x}-{\bf
x'}),\nonumber\\
\{\psi_\alpha({\bf x},t),\psi_\beta({\bf x'},t)\}&=&0.
\end{eqnarray}
The Hamiltonian of the total system can also be find as
\begin{eqnarray} \label{d3}
&&H =c \int {d^3 {\bf x}\,}  \psi
^\dag ({\bf x},t) \left[{{\boldsymbol{ \alpha }} \cdot\left({{\bf p}-{\bf R}(t)-e{\bf A}({\bf x},t)}\right) + (mc^2+V({\bf x}))  \beta }\right]\psi ({\bf x},t)
\nonumber \\
&&+\frac{1}{2} \int_0^\infty d\omega ({\bf Q}^2 (\omega,t)+ \omega
^2 {\bf X}^2(\omega,t))+\frac{1}{8\pi\epsilon_0}\int {d^3 {\bf
x}\,}\int {d^3 {\bf x'}\,}\frac{\rho({\bf x},t)\rho({\bf
x'},t)}{|{\rm x}-{\rm x'}|} \nonumber\\
\end{eqnarray}
where ${\bf p}=-\imath\hbar\nabla$, the field ${\bf R}(t)$ is defined by Eq.~(\ref{a7}) and $\rho({\bf
x},t)=e\psi^\dag ({\bf x},t)\psi ({\bf x},t)$ is the charge density. In the Heisenberg picture, by using commutation relations (\ref{d2/2}) and the total Hamiltonian (\ref{d3}), the relativistic analogues of the motion equation (\ref{a6}) is obtained as
\begin{equation} \label{d4}
\ddot { X}_i(\omega,t) + \omega ^2  { X}_i(\omega,t) = {
f}_{ji}(\omega ){ J}_j(t),
\end{equation}
with the solution
\begin{eqnarray} \label{d5}
{ X}_i (\omega,t )  =  Q_i (\omega,0 ) \frac{{\sin \omega
t}}{\omega } +
 {\rm X}_i(\omega,0 )\cos \omega t + \int_0^t  {dt'}
 \frac{{\sin \omega (t - t')}}{\omega }
 { f}_{ji}(\omega){J}_j(t^\prime),\nonumber\\
\end{eqnarray}
where
\begin{eqnarray}\label{d6}
&&{\bf J}(t) = c\int {d^3 {\bf x}\,} \psi ^\dag ({\bf
x},t){\boldsymbol {\alpha} }\,\psi ({\bf x},t).
\end{eqnarray}
Now by substituting ${X}_i (\omega,t)$ from (\ref{d5}) in (\ref{a7}) we find
\begin{equation}\label{d7}
{{ R}_i}(t) = \int_0^\infty  {dt\,} \chi_{ij}(t - t')J_j(t')\, + {
R}_{i}^{\rm N}(t),
\end{equation}
where $\chi_{ij}$ and ${R}_{i}^{(N)}(t)$ are the causal susceptibility tensor of the medium and the noise operator that perviously defined by Eqs.~(\ref{a10}) and (\ref{a11}), respectively. If we apply the Heisenberg equation to the
Dirac field $\psi ({\bf x},t)$ and make use of the anticommutation relations (\ref{d2/2}), then Dirac
equation in the presence of a dissipative medium are found as
\begin{equation}\label{d8}
\iota \hbar \dot \psi ({\bf x},t) =  \left[{ c\,{\boldsymbol{ \alpha }} \cdot
{\boldsymbol\pi}+e\phi({\bf x},t)+(mc^2+V({\bf x}) )\beta} \right]\psi ({\bf x},t),
\end{equation}
in which ${\boldsymbol\pi}\equiv\left({ {\bf p}-{\bf R}(t)-e{\bf A}({\bf x},t)} \right)$, and
\begin{equation}\label{d8/1}
\phi({\bf x},t)=\frac{e}{4\pi\epsilon_0}\int {d^3 {\bf
x'}\,}\frac{\psi({\bf x'},t)\psi({\bf x'},t)}{|{\bf x}-{\bf x'}|}.
\end{equation}
Substitution Eq.~(\ref{d7}) in (\ref{d8}), a relativistic Langevin equation for a Dirac particle in an
anisotropic-dissipative-medium can be obtained as
\begin{eqnarray}\label{d9}
&&\iota \hbar \dot \psi ({\bf x},t) + c{\boldsymbol{ \alpha }}
\cdot( \imath\hbar\nabla+e{\bf A}({\bf x},t))\psi({\bf
x},t)+\int_0^\infty dt'\,
\alpha_i\,\chi_{ij}(t-t')\, J_j(t')\,\,\psi({\bf x},t)\nonumber\\
&&-[e\phi({\bf x},t)+(mc^2+V({\bf x})) \beta ] \psi ({\bf
x},t)=-{\boldsymbol{ \alpha }} \cdot {\bf R}^{\rm N}(t)\,\,\psi({\bf
x},t).
\end{eqnarray}
What we have here is a case of the fluctuation-dissipation relation.
Generally speaking, if a system is coupled to a dissipative medium that can take energy
from the system in an effectively irreversible way, then the medium must also
cause fluctuations. The fluctuations and the dissipation effects go hand in hand, we
cannot have one without the other. In Eq.~(\ref{d9}), the effects of the dissipation and the fluctuation are appeared as two contributions $\chi_{ij}(t-t')$  and $ {\bf R}^{\rm N}(t)$ from the vector field ${\bf R}$. Thus, our approach suggests the simplest way in which the dissipation and the fluctuation effects can emerge from the classical to the relativistic quantum regime. In fact, based on a minimal coupling scheme, we can introduce the dissipation and the fluctuation effects of the medium by writing ${\bf p}^2$ equivalently as $({\bf p}-{\bf {R}})^2$ in the Heisenberg equation for a free brownian particle, or making the replacement $({\bf p}-{\bf {R}})^2 \rightarrow ({\bf p}-{\bf {R}}-e{\bf A})^2$ in presence of the electromagnetic field. This approach to the derivation of the motion equation is simpler than that proceeding from other approach~\cite{47,48} and provides a consistent and rigorous basis for the introduction of the dissipation and the fluctuation in a Hamiltonian formalism and motion equations from the classical to the relativistic domain.

Let us examine the nonrelativistic limit of the Dirac equation~(\ref{d8}). we introduce the Dirac field $\psi({\bf
x},t)$ as
\begin{equation}\label{d8/2}
\psi= \left( {\begin{array}{*{20}c}
   \tilde{\eta}  \\
   \tilde{\xi}  \\
\end{array}} \right),
\end{equation}
where $\tilde{\eta}$ and $\tilde{\xi}$ are each two-component column vectors, so that by applying Eq.~(\ref{d8}) we have
\begin{equation}\label{d8/3}
\imath\hbar\frac{\partial}{\partial t}\left( {\begin{array}{*{20}c}
   \tilde{\eta}  \\
   \tilde{\xi}  \\
\end{array}} \right)=c\,{\boldsymbol{\sigma}} \cdot
{\boldsymbol\pi}\left( {\begin{array}{*{20}c}
   \tilde{\xi}  \\
   \tilde{\eta}  \\
\end{array}} \right)+e\phi({\bf x},t)\left( {\begin{array}{*{20}c}
   \tilde{\eta}  \\
   \tilde{\xi}  \\
\end{array}} \right)+(mc^2+V({\bf x}) )\left( {\begin{array}{*{20}c}
   \tilde{\eta}  \\
   -\tilde{\xi}  \\
\end{array}} \right).
\end{equation}
In the nonrelativistic limit the energy $mc^2$ is large compared with any
kinetic or potential energy, and this suggests writing
\begin{equation}\label{d8/4}
 \left( {\begin{array}{*{20}c}
   \tilde{\eta}  \\
   \tilde{\xi}  \\
\end{array}} \right)= e^{-\imath\frac{ mc^2 }{\hbar}\,t}\left( {\begin{array}{*{20}c}
   {\eta}  \\
   {\xi}  \\
\end{array}} \right),
\end{equation}
in which assuming that $\eta$ and $\xi$ slowly varying compared with $e^{-\imath\frac{ mc^2 }{\hbar}\,t}$ in a
nonrelativistic approximation. Therefore, Eq.~(\ref{d8/3}) becomes
\begin{equation}\label{d8/5}
\imath\hbar\frac{\partial}{\partial t}\left( {\begin{array}{*{20}c}
  {\eta}  \\
  {\xi}  \\
\end{array}} \right)=c\,{\boldsymbol{\sigma}} \cdot
{\boldsymbol\pi}\left( {\begin{array}{*{20}c}
  {\xi}  \\
   {\eta}  \\
\end{array}} \right)+e\phi({\bf x},t)\left( {\begin{array}{*{20}c}
   {\eta}  \\
   {\xi}  \\
\end{array}} \right)-2mc^2 \left( {\begin{array}{*{20}c}
   {0}  \\
   {\xi}  \\
\end{array}} \right)+V({\bf x}) \left( {\begin{array}{*{20}c}
   {\eta}  \\
   -{\xi}  \\
\end{array}} \right),
\end{equation}
and in the nonrelativistic limit the second of the two indicated equations is
replaced by
\begin{equation}\label{d8/5}
{\xi}\approx \frac{\,{\boldsymbol{\sigma}} \cdot
{\boldsymbol\pi}}{2mc}\eta.
\end{equation}
Then the equation for $\eta$ becomes
\begin{equation}\label{d8/5}
\imath\hbar\frac{\partial \eta}{\partial t}\approx \left[{ \frac{({\boldsymbol{\sigma}} \cdot
{\boldsymbol\pi)^2}}{2m}+ e\phi({\bf x},t)+V({\bf x})}\right]\eta.
\end{equation}
This result can be cast in a more familiar form by using the general
identity $({\boldsymbol{\sigma}} \cdot {\bf C})({\boldsymbol{\sigma}} \cdot {\bf D})={\bf C}\cdot{\bf D}+\imath {\boldsymbol{\sigma}} \cdot ({\bf C}\times{\bf D})$~\cite{68}, and we can write~(\ref{d8/5}) for a spinless particle as the nonrelativistic equation
\begin{equation}\label{d8/6}
\imath\hbar\frac{\partial \eta}{\partial t}= \left[{ \frac{\left({ {\bf p}-{\bf R}(t)-e{\bf A}({\bf x},t)} \right)^2}{2m}+ e\phi({\bf x},t)+V({\bf x})}\right]\eta.
\end{equation}
Comparison above obtained result with Eq.~(\ref{d8}) suggests the interpretation of ${\boldsymbol{\alpha}}$ as the operator corresponding to the particle's velocity, i.e., $c\int d^3{\bf x}\,\psi^\dag({\bf x},t)\,{\boldsymbol{ \alpha }}\,\psi ({\bf x},t)={\bf \dot{q}}(t)$. This interpretation is strengthened by the Heisenberg equation ${\bf q}$ that follow straightforwardly from the Hamiltonian~(\ref{d3}) and the canonical commutation relations~(\ref{b2}):
\begin{eqnarray} \label{b7}
&& {\bf \dot q}(t) = \frac{\iota }{\hbar }\left[ {H,{\bf q}(t)}
\right] = c\int d^3{\bf x}\psi^\dag({\bf x},t){\boldsymbol{ \alpha }}\psi ({\bf x},t).
\end{eqnarray}

Let us expand the Dirac field $\psi({\bf x},t)$ in term of the eigenfunctions of the free Dirac equation in the
absence of dissipative medium
\begin{equation}\label{d10}
\psi ({\bf x},t) = \frac{1}{{(2\pi )^{{3 \mathord{\left/ {\vphantom
{3 2}} \right. \kern-\nulldelimiterspace} 2}} }}\sum\limits_{\mu  =
1}^4 \int {d^3 {\bf q}\,} c_\mu  ({\bf q},t)\psi _\mu  ({\bf q}),
\end{equation}
where $\psi _\mu  ({\bf q}) = u_\mu  ({\bf q})e^{\iota {\bf q} \cdot {\bf x}} $ and $u_\mu  ({\bf q})$ are
four-component spinors of the Dirac equation with corresponding eigenvalues $E_{{\bf q}} = \pm \sqrt {\hbar ^2 c^2
{\bf q}^2 + m^2 c^4 } $ and normalization $u^\dag_\mu({\bf q}) u_\nu({\bf q})=\delta_{\mu\nu}$ \cite{53}. Here, the operator $c_\mu ({\bf q}\,,t)$ annihilates a particle with momentum $\hbar{\bf q}$. Substitution Eq.~(\ref{d10}) in
(\ref{d3}), the Hamiltonian of the total system can be written as
\begin{equation} \label{d11}
H=\sum\limits_{\mu  = 1}^4 {} \int {d^3 {\bf q}\,} E_{{\bf q} } c_\mu ^\dag ({\bf q}\,,t)c_\mu  ({\bf
q}\,,t)+H_{\rm m}+H_{\rm F} +H_{\rm int},
\end{equation}
where
\begin{eqnarray}
H_{\rm int}&=&-c\sum\limits_{\mu,\mu' = 1}^4 \int_0^\infty {d\omega } \int {d^3 {\bf q}} \sqrt {\frac{{\hbar
}}{{2\omega}}}\,u_\mu^\dag({\bf q}) \alpha_i\, f_{ij}(\omega )u_{\mu'}({\bf q})\{ c_\mu ^\dag ({\bf q})c_{\mu'}
({\bf q} )b_j
(\omega )\nonumber\\
&&+h.c.\}\nonumber\\
&&- e\sum_{\lambda = 1}^2\sum\limits_{\mu,\mu' = 1}^4  \int {d^3 {\bf k}}\int {d^3 {\bf q}} \sqrt {\frac{{\hbar
c }}{{2(2\pi)^3\epsilon_0|{\bf k}|}}}\,\{\,u_{\mu'}^\dag({\bf k}+{\bf q}) \boldsymbol{\alpha}\cdot
e_\lambda({\bf k})u_{\mu}({\bf
q})\,\nonumber\\
&&\hspace{4cm} \times c_{\mu'}^\dag ({\bf k}+{\bf q})c_\mu  ({\bf q} )a_\lambda ({\bf k})+h.c.\},
\end{eqnarray}
and $H_{\rm m}$ and $H_{\rm F}$ are defined in Eqs. (\ref{b24}) and (\ref{f12}), respectively. The Heisenberg equation for the operator $a_\lambda(\bf k,t)$ is found from Hamiltonian (\ref{d11}) as
\begin{eqnarray}\label{d12}
\dot{a}_\lambda({\bf k})&=&\frac{\imath}{\hbar}[H,a_\lambda({\bf
k})]=-\imath \omega_{\bf k}a_\lambda({\bf
k})\nonumber\\
&+& {\frac{{ \imath e c }}{\sqrt{2(2\pi)^3\hbar\epsilon_0\omega_{\bf
k}}}}\sum_{\mu,\mu'}\int {d^3 {\bf q}} (\,u^\dag_{\mu}({\bf
q})\boldsymbol{\alpha}\cdot e_\lambda({\bf k})u_{\mu'}({\bf k}+{\bf
q}) \,)c^\dag_\mu ({\bf q} )c_{\mu'} ({\bf k}+{\bf q}),\nonumber \\
\end{eqnarray}
with the formal solution
\begin{eqnarray}\label{d13}
a_\lambda  ({\bf k},t)&=& e^{ - \iota \omega _{\bf k} t} a_\lambda
({\bf k},0) \nonumber\\
&&+ {\frac{{ \imath e c }}{\sqrt{2(2\pi)^3\hbar\epsilon_0\omega_{\bf
k}}}}\sum_{\mu,\mu'}\int {d^3 {\bf q}}\,C_{\lambda,\mu,\mu'} ({\bf
k},{\bf q})\nonumber\\
&&\times\int_0^t {dt'} \,e^{ - \iota \omega _{\bf k} (t -
t')}c^\dag_\mu
({\bf q}\,,t' )c_{\mu'} ({\bf k}+{\bf q}\,,t')\nonumber\\
&=&a_{0,\lambda} ({\bf k},t)+a_{RR,\lambda} ({\bf k},t),
\end{eqnarray}
where $C_{\lambda,\mu,\mu'}({\bf k},{\bf q})\equiv u^\dag_{\mu}({\bf q})\boldsymbol{\alpha}\cdot {\bf e}_\lambda({\bf
k})u_{\mu'}({\bf k}+{\bf q})$. Now, by substitution Eq. (\ref{d13}) into (\ref{d9}) a relativistic-Langevin-equation is obtained that describes a relativistic moving particle through an anisotropic-dissipative-medium in presence of the electromagnetic field
\begin{eqnarray}\label{d14}
&&\iota \hbar \dot \psi ({\bf x},t) + c\,{\boldsymbol{ \alpha }}
\cdot( \imath\hbar\nabla+e{\bf A}_{0}({\bf x},t))\psi({\bf
x},t)-[e\phi({\bf x},t)+(mc^2+V({\bf x})) \beta ] \psi ({\bf
x},t)\nonumber\\
&&+\int_0^\infty dt'\,
\alpha_i\,\chi_{ij}(t-t')\, J_j(t')\,\,\psi({\bf x},t)=-{\boldsymbol{ \alpha }} \cdot ({\bf R}^{\rm N}(t)+ec{\bf
A}_{RR}({\bf x},t))\,\,\psi({\bf x},t),\nonumber \\
\end{eqnarray}
where
\begin{equation}\label{d15}
{\bf A}_{0}({\bf r},t) = \sum\limits_{\lambda  = 1}^2 {\int
{d^3 {\bf k}} \sqrt {\frac{\hbar }{{2(2\pi )^3 \varepsilon _0
\omega_{\bf k} }}} } \left( {a_{0,\lambda} ({\bf k},t)e^{\iota {\bf
k} \cdot {\bf r}} + h.c. } \right){\bf e}_\lambda  ({\bf k}),
\end{equation}
and
\begin{eqnarray} \label{d16}
{\bf A}_{RR}({\bf r},t) = \sum\limits_{\lambda  = 1}^2 {\int
{d^3 {\bf k}} \sqrt {\frac{\hbar }{{2(2\pi )^3 \varepsilon _0
\omega_{\bf k} }}} } \left( {a_{RR, \lambda } ({\bf k},t)e^{\iota {\bf
k} \cdot {\bf r}} + h.c. } \right){\bf e}_\lambda  ({\bf k}),
\end{eqnarray}
are the vacuum and the radiation reaction contributions, respectively. It is seen that, the coupling of the dirac field to the electromagnetic field has a dissipative component, in the form of radiation reaction, and a fluctuation component, in the form of the vacuum field. Given the existence of radiation reaction, the vacuum field must also exist in order to satisfy the fluctuation-dissipation relation~\cite{68}.\\
The vacuum field and the radiation-reaction field can be calculated from the time derivative of Eqs.~(\ref{d15}) and~(\ref{d16}). It is interesting to compare these latter fields with the result of the nonrelativistic quantum mechanic~(\ref{f17}) and (\ref{f18}). At first glance, we can easily find that the vacuum field contributions are the same on the one hand. On the other hand, in nonrelativistic theory, the "velocity" is $c\int d^3{\bf x}\psi^\dag({\bf x},t){\boldsymbol{ \alpha }}\psi ({\bf x},t)\rightarrow{\bf \dot{q}}(t)$ and therefore, by using Eqs.~(\ref{d13}) and (\ref{d10}) we have
\begin{eqnarray}\label{d17}
c\, {\bf e}_\lambda({\bf
k})\cdot \sum_{\mu,\mu'}\int {d^3 {\bf q}}u^\dag_{\mu}({\bf q})\boldsymbol{\alpha}u_{\mu'}({\bf k}+{\bf q})c^\dag_\mu
({\bf q}\,,t' )c_{\mu'} ({\bf k}+{\bf q}\,,t')\nonumber\\=c\, {\bf e}_\lambda({\bf k})\cdot\int d^3{\bf x}\psi^\dag ({\bf x},t){\boldsymbol{ \alpha }}\psi ({\bf x},t)\rightarrow {\bf e}_\lambda({\bf k})\cdot{\bf \dot{q}}(t).
\end{eqnarray}
Substitution Eq.~(\ref{d17}) into the time derivative of Eq.~(\ref{d16}), we indeed revert to the nonrelativistic expression (\ref{f18}) for the radiation reaction field.

%
\subsection{Relativistic quantum theory of Cherenkov radiation}
In order to illustrate the applicability of our approach, we attempt to treat the relativistic theory
of Cherenkov radiation in the presence of an anisotropic polarizable medium. Cherenkov radiation is the radiation with continuous spectrum that emitted by the medium due to the motion of a charged particle moving through the medium with a velocity exceeding the phase velocity of light in it. We consider a charge particle with mass $m$ and electric charge $e$ uniformly moving in the anisotropic polarizable medium which describe by the Hamiltonian~(\ref{d11}) with the field operator $\bf R$ that now plays the role of the polarization density of the medium. The Hamiltonian operator of the total system~(\ref{d11}), i.e. the electromagnetic field, the polarizable medium and the particle, in the large-time limit when the medium and electromagnetic field tend to an equilibrium state can be rewritten as follows~\cite{70}
\begin{eqnarray} \label{t1}
H&=&H_0+H_{int},
\end{eqnarray}
where
\begin{equation}\label{t2}
H_0=H_{ele}+H_F,
\end{equation}
with
\begin{equation}\label{t3}
H_{ele} = \sum\limits_{\mu  = 1}^4 {} \int {d^3 {\bf q}\,} E_{{\bf
q} }\, c_\mu ^\dag ({\bf q},t)c_\mu  ({\bf q}\,,t),
\end{equation}
\begin{equation}\label{t4}
H_F = : \int d\omega\hbar\omega\int d^3 {\bf
k}\, b_i ^\dag ({\bf k},\omega,t)b_i({\bf
k},\omega,t)\}:,\nonumber\\
\end{equation}
and
\begin{eqnarray}\label{t5}
&&H_{int}=\iota ce
\sum\limits_{\mu ,\mu ' = 1}^4  \int {d^3 {\bf k}} \int {d^3 {\bf
q}} \int_0^\infty  {d\omega } \sqrt {\frac{{\hbar \omega }}{{2(2\pi
)^3\varepsilon_0 }}}\{u_\mu ^\dag({\bf q}){\alpha}_i \nonumber\\
&&\times {\rm{\textsf{G}}}_{ij}({\bf k},\omega)f_{jl}(\omega) u_{\mu'} ({\bf q} - {\bf k}) c_\mu ^\dag ({\bf
q})c_{\mu '} ({\bf q} - {\bf k})b_l  {\bf (k},\omega
,0)e^{-\imath\omega t}-h.c.\}. \nonumber\\
\end{eqnarray}
Here, the different components of the Green tensor~${\rm{ \textsf{G}}}_{jl}({\bf k},\omega)$ in Eq.~(\ref{t5}) satisfy the following set of algebraic equations
\begin{equation}\label{t6}
{\rm{\textsf{H}}}_{ij}({\bf k},\omega){\rm{ \textsf{G}}}_{jl}({\bf k},\omega)=\delta_{il},
\end{equation}
where
\begin{equation}\label{t7}
{\rm{\textsf{H}}}_{ij}({\bf k},\omega)=(\mu_0^{-1}k^2c^2-\omega^2\epsilon_{ij}(\omega)),
\end{equation}
in which the permittivity tensor $\epsilon_{ij}(\omega)$ is related to the susceptibility~(\ref{a12}) as $\epsilon_{ij}(\omega)=1+\chi_{ij}(\omega)$. It is seen from Eq.~({\ref{t6}}), looked upon either as a matrix or as an operator equation, that the response tensor~${\rm{ G}}_{ij}({\bf k},\omega)$ is the
inverse of~${\rm{ H}}_{ij}({\bf k},\omega)$. The solution of it is obtained most easily by evaluating
the inverse of~${\rm{ H}}_{jl}({\bf k},\omega)$ with the help of dyadic analysis~\cite{72}.

The unperturbed Hamiltonian $H_0=H_{ele}+H_F$ has the eigenstate $\mid ele+ rad\rangle=\mid ele\rangle \otimes\mid rad\rangle$ which are the direct product of the eigenstates of $H_{ele}$ and
$H_F$. We apply the perturbation theory up to the first order to treat the
transition probability per unit time for a free Dirac particle of
momentum $\hbar {\bf q}$ to emit a photon of momentum $\hbar {\bf
k}$ and energy $\hbar\omega$, thereby changing its momentum to $\hbar
({\bf q}-{\bf k})$
\begin{eqnarray}\label{t9}
&&\Gamma _{{\bf q} \to {\bf q} - {\rm k}} = \frac{{2\pi }}{\hbar
}\left| \langle 1_{\bf k} \mid \langle {\bf q}-{\bf k} \mid H_{int
}\mid {\bf q}\rangle \mid  0\rangle \right|^2\nonumber\\
&& \times \delta  \left( { \sqrt {\hbar
^2 c^2 {\bf q}^2 + m^2 c^4 } - \sqrt {\hbar ^2 c^2 |{{\bf q} - {\bf
k}}|^2 + m^2 c^4 } - \hbar\omega } \right),
\end{eqnarray}
where the states $\mid  0\rangle$ and $ \mid  1_{\bf k}\rangle$ present the vacuum state
of the electromagnetic field and the excited state of the
electromagnetic field with a single photon with wave vector
${\bf k}$ and frequency $\omega$, respectively. The argument of the Dirac $\delta$ function displays the
conservation of energy. The radiation intensity in a form of the Cherenkov radiation is obtained by multiplying Eq.~(\ref{t9})
by $\hbar\omega$ and integrating over ${\bf k}$ and $\omega$. We find that
\begin{equation}\label{t10}
\frac{dW}{dt}= \frac{1}{2}\sum\limits_{\lambda  = 1}^2 \sum\limits_{\mu ,\mu ' = 1}^2\int d^3{\bf
k}\int_0^\infty d\omega \hbar\omega \Gamma _{{\bf q} \to {\bf q} - {\bf k}}.
\end{equation}
Here, the sum is taken over the final spin states of the particle with positive energy,
$(\mu=1,2)$, as well as an average over the initial spin states.
In order to calculate above equation we need to evaluate the following sum
\begin{eqnarray}\label{t11}
S = \frac{1}{2}\sum\limits_{\lambda  = 1}^2  \sum\limits_{\mu ,\mu ' = 1}^2 \left| {u_\mu ^\dag ({\bf q})\,{\boldsymbol{\alpha}}\cdot  {\bf{\textsf{G}}}({\bf k},\omega)\cdot {\bf f}(\omega)\,u_{\mu '} ({\bf q} - {\bf k})} \right|^2.
\end{eqnarray}
For this purpose, we introduce the annihilation operator \cite {70}
\begin{eqnarray}\label{t12}
&&\Lambda ({\bf q}) = \frac{{c\boldsymbol{\alpha} \cdot {\bf q} + \beta mc^2  + \left| {E_{\bf q} } \right|}}{{2\left|
{E_{\bf q} }
\right|}}.
\end{eqnarray}
It is straightforward with the help of Eq.~(\ref{t12}) to show that
\begin{eqnarray}\label{t13}
S&=&\frac{1}{8}{\rm Tr}[({\boldsymbol{\alpha}}\cdot  {\bf{\textsf{G}}}({\bf k},\omega)\cdot {\bf f}(\omega))\Lambda ({\bf q} - {\bf k})({\boldsymbol{\alpha}}\cdot  {\bf{\textsf{G}}}({\bf k},\omega)\cdot {\bf f}(\omega))^\dag\Lambda ({\bf q})]\nonumber\\
&=&  { v}_i \,{\rm{Im}}[{\rm{\textsf{G}}}^\bot_{ij}({\bf k},\omega)]\, { v}_j + \frac{1}{2}\{ 1 - \sqrt {(1 - {{{\bf v}^2 }
\mathord{\left/ {\vphantom {{{\rm v}^2 } {c^2 }}} \right. \kern-\nulldelimiterspace} {c^2 }})(1 - {{{\bf
v'}^2 } \mathord{\left/ {\vphantom {{{\rm v}_2^2 } {c^2 }}} \right. \kern-\nulldelimiterspace} {c^2 }})}  -
\frac{{{\bf v}  \cdot {\bf v'} }}{{c^2 }}\}.
\end{eqnarray}
in which ${\bf v} = {{\hbar c^2 {\bf q}} \mathord{\left/ {\vphantom {{c^2 {\bf q}} {E_{\bf q} }}} \right. \kern-\nulldelimiterspace} {E_{\bf q} }}$ and ${\bf v'} = {{\hbar c^2 {\bf q'}} \mathord{\left/ {\vphantom {{c^2 {\bf q'}} {E_{\bf q} }}} \right. \kern-\nulldelimiterspace} {E_{\bf q'} }}$ are the velocities before an after the emission of the photon, respectively.
Notice that in writing Eq.~(\ref{t13}), we have used the tensor identity
\begin{equation}\label{t13/1}
\frac{\omega^2}{c^2}{\bf{\textsf{G}}}({\bf k},\omega)\cdot{\rm{Im}}[{\boldsymbol\chi}(\omega) ] \cdot {\bf{\textsf{G}}}^*({\bf k},\omega)={\rm{Im}}[{\bf{\textsf{G}}}({\bf k},\omega)],
\end{equation}
as well as just the transverse part of the Green tensor are considered in calculation, since the transverse contribution yields the radiative effect, while the longitudinal contribution provides purely nonradiative effects. On the other hand, the second term in Eq.~(\ref{t13}) has no contribution to the radiation intensity~(\ref{t10}), since the wave length of the electron is much smaller than the photon's wavelength. Using these notes one can easily show that
\begin{eqnarray}\label{t14}
\frac{dW}{dt} &=& \frac{{e^2}}{{4\pi^3 \varepsilon _0 }} \int_0^{
+ \infty }  d^3{\bf k}\int_0^{ + \infty } \,d\omega\, \omega\, { v}_i \,{\rm{Im}}[{\rm{G}}^\bot_{ij}({\bf k},\omega)]\, { v}_j \nonumber\\
&&\times\delta\left({{\bf v}\cdot{\bf k}-\omega[1 + \frac{{\hbar \omega }}{{2mc^2 }}(\frac{{{ k}^2 c^2 }}{{\omega ^2 }} - 1)\sqrt {1
- \frac{{{\bf v}^2 }}{{c^2 }}} ]}\right).
\end{eqnarray}
Now, it is interesting to compare this recent relation with the result of the relativistic Cherenkov radiation in presence of isotropic medium~\cite{70}. In the latter, the transverse Green tensor~(\ref{t6}) is given by
\begin{equation}\label{t15}
{\rm{\textsf{G}}}^\bot_{ij}({\bf k},\omega)=\frac{\delta_{ij}-k_ik_j/k^2}{{ \mu_0^{-1}k^2 c^2 - \omega ^2 \varepsilon (\omega ) }}.
\end{equation}
Let $\theta$ be the angle between ${\bf q}$ and ${\bf k}$, according to the argument of the Dirac $\delta$ function~({\ref{t9}}), the photon is emitted at an angle to the path of the particle as
\begin{equation}\label{t16}
\cos\theta=\frac{\omega }{{vk}}\left({1 + \frac{{\hbar \omega }}{{2mc^2 }}(\frac{{{ k}^2 c^2 }}{{\omega ^2 }} - 1)\sqrt {1
- \frac{{{\bf v}^2 }}{{c^2 }}}}\right).
\end{equation}
Substitution of Eqs.~(\ref{t15}) and~(\ref{t16}) into Eq.~({\ref{t14}}) and integration over the polar angel $\theta$ and the azimutal angel $\varphi$, yields that in the presence of isotropic medium
\begin{eqnarray}\label{t17}
\frac{dW}{dt} &=& \frac{{e^2 v}}{{2\pi ^2 \varepsilon _0 }} \int_0^{
+ \infty }  dkk\int_0^{ + \infty } \,d\omega \omega \,\left( {1 -
\frac{\omega^2 }{{v^2k^2}}[1 + \frac{{\hbar \omega }}{{2mc^2
}}(\frac{{k^2 c^2 }}{{\omega ^2 }} -
1)\sqrt {1 - \frac{{v^2 }}{{c^2 }}} ]^2} \right)\nonumber\\
&\times &{\mathop{\rm Im}\nolimits} \left( {\frac{1}{{  \mu_0^{-1}k^2 c^2 - \omega ^2 \varepsilon (\omega )}}} \right),
\end{eqnarray}
which is consistent with the result has been reported in \cite{70}.
%
%
\section{Conclusion}
In this paper, the classical, the non-relativistic and the relativistic quantum dynamics of a Brownian particle under the influence of an arbitrary potential in anisotropic medium is canonically investigated. In this formalism the dissipative medium is modeled by the collections of harmonic oscillators. A fully canonical
quantization of the dynamical variables that modeling the dissipative medium is demonstrated and, as a direct consequence of it the susceptibility tensor of the dissipative medium in terms of the coupling functions are calculated. In Heisenberg picture, a non-relativistic and relativistic quantum Langevin-equation is derived and explicit expressions for quantum noise and dynamical variables of the system are obtained. It is shown that how radiation reaction in this theory can be expressed in terms of the source operators, then the radiated spectrum of the medium as the cherenkov radiation is calculated. We finally introduced a minimal coupling scheme to enter the dissipation and the fluctuation of the medium in the Hamiltonian formalism and the motion equations from the classical to the relativistic regime.
\\
\\
\textbf{Acknowledgments}\\
E. Amooghorban wish to thank the Shahrekord University for their support.

\appendix
\section{}
In this appendix we evaluate the time-dependent coefficient $c(t)$ in Eq. (\ref{c4}) for the spontaneous decay
of an initially excited atom embedded in anisotropic dissipative medium. By substituting $\left|{\psi
(t)}\right\rangle $ from Eq.(\ref{c4}) into Eq.(\ref{c5}) and using the expansions (\ref{a7}), (\ref{b21}) and
(\ref{b24}), we find the following coupled differential equations
\begin{eqnarray}\label{e1}
&&\dot{c}(t)=-\imath\omega _0 \, c(t) -
\int_0^\infty {d\omega } \sqrt {\frac{\hbar\omega }{{2
}}}{ q}_{12i}^* \,f_{ij}(\omega )M_j (\omega,t),\\
&&\dot M_i (\omega ,t) = -\imath\omega \,M_i (\omega ,t)
+ \sqrt {\frac{{ \omega }}{2\hbar^3}}{ q}_{12j}\,
f_{ji}(\omega )\, \,c(t).
\end{eqnarray}
We can solve these coupled differential equations by using Laplace transformation technique. Let $\tilde{c}(s)$
denotes the Laplace transform of $c(t)$. Taking the Laplace transform of Eqs. (\ref{e1}), combining them and
using the relations (\ref{b12}), we find
\begin{equation}\label{e2}
s\tilde c(s) = c(0)-\imath\omega _0 \,\tilde c(s)
+\frac{1}{\hbar} [{\rm q}_{\,12,i}^* \widetilde{G}_{ij} (\imath s){q}_{12,j}]\tilde c(s),
\end{equation}
where
\begin{equation}\label{e3}
\widetilde{G}_{ij}(\imath s) = \int_0^\infty  {d\omega } \frac{{\omega ^2
}}{{\pi (\imath s - \omega
)}}{\rm Im}\,\chi_{ij} (\omega).
\end{equation}
The tensor $\widetilde{G}_{ij}(\imath s)$ gives the spontaneous-emission and frequency-shift of the atom due to the
presence of dissipative medium. From definition (\ref{e3}) it is obvious that $\widetilde{G}_{ij}(\imath s)$ is an
analytic tensor in the upper half-plane $Re(s)> 0 $, therefore
\begin{equation}\label{e4}
\dot c(t) =  - \imath\omega _0 c(t) + \int_0^t {K (t -
t')\,c(t')\,dt'},
\end{equation}
where
\begin{equation}\label{e5}
K (t - t') = \frac{1}{2\pi\imath\hbar}\int_{-\infty }^{+\infty} {du
\,e^{-\imath u(t - t')} [{ q}_{\,12,i}^* \widetilde{G}_{ij}(u
+\imath \tau){ q}_{\,12,j}}].
\end{equation}
Here we use the Markov's approximation \cite{68} and replace $ c(t')$ in (\ref{e4}) by
\begin{equation}\label{e6}
c(t') = c(t)e^{\imath\omega_0 (t - t')}.
\end{equation}
By lengthy but straightforward calculations and using Kramers-Kroning relations we deduce
\begin{equation}\label{e7}
\dot c(t) =  - \imath\omega _0 c(t) - (\,\,\Gamma + \imath\Delta
)\,c(t)
\end{equation}
where
\begin{equation}\label{e8}
\Gamma =  -\frac{1}{\hbar}\left[{{\rm q}_{12,i}^*\,
{\rm Im}[\widetilde{G}_{ij}(\omega_0+\imath 0^+ )]\,{\rm q}_{12,j}} \right]=\frac{\omega_0^2}{\hbar}\left[{{\rm q}_{12,i}^*\,
{\rm Im}[{ \chi}_{ij}(\omega_0 )]\,{\rm q}_{12,j}} \right],
\end{equation}
and
\begin{equation}\label{e9}
\Delta =  -\frac{1}{\hbar}\left[{{\rm q}_{12,i}^*\,
{\rm Re}[\widetilde{G}_{ij}(\omega_0+\imath 0^+ )]\,{\rm q}_{12,j}} \right]=P\int_0^\infty {d\omega } \frac{{q}_{12,i}^*\,
{\rm Im}[{ \chi}_{ij}(\omega )]\,{\rm q}_{12,j}}{{\pi\hbar(\omega_0 - \omega )}}.
\end{equation}

\end{document}